\documentclass[aps,pra,reprint,superscriptaddress,amsmath,amssymb,floatfix]{revtex4-2}

\usepackage{graphicx}
\usepackage{bm}
\usepackage{braket}
\usepackage{algorithm}
\usepackage{algpseudocode}
\usepackage{amsthm}
\usepackage{booktabs}
\usepackage[svgnames]{xcolor}
\usepackage{hyperref}
\hypersetup{
     colorlinks = true,
     linkcolor = Maroon,
     anchorcolor = black,
     citecolor = DarkBlue,
     filecolor = blue,
     urlcolor = DarkBlue
     }

\newcommand{\Tr}{\operatorname{Tr}}
\newcommand{\tr}{\operatorname{tr}}
\newcommand{\ii}{\mathrm{i}}
\newcommand{\E}{\mathbb{E}}
\newcommand{\Cov}{\operatorname{Cov}}
\newcommand{\id}{\mathbb{I}}
\newtheorem{theorem}{Theorem}
\newtheorem{proposition}{Proposition}
\newtheorem{lemma}{Lemma}

\begin{document}


\title{Complex quasiprobabilities as measurable temporal interference}

\author{Wenzheng Dong}
\affiliation{School of Physical and Chemical Sciences, Queen Mary University of London, London E1 4NS, United Kingdom}

\author{Alexander Yosifov}
\affiliation{Clarendon Laboratory, University of Oxford, Parks Road, Oxford OX1 3PU, United Kingdom}

\author{Jinzhao Sun}
\affiliation{School of Physical and Chemical Sciences, Queen Mary University of London, London E1 4NS, United Kingdom}

\begin{abstract}
Quantum dynamics can encode interference information that conventional correlation functions leave unresolved. Here we show that the imaginary component of the extended
Kirkwood-Dirac quasiprobability reveals dynamical distinctions absent from out-of-time-order correlators (OTOCs) formed from two repeated Hermitian observables at infinite temperature. We interpret its real and imaginary components as complementary quadratures of interference between many-body echo amplitudes. We prove that the imaginary component vanishes for every unitary
evolution if either projective measurement has two outcomes, whereas nonreal values arise for generic unitary dynamics when it has at least three (e.g., spin-1 systems). We develop an algorithm to reconstruct the full quasiprobability from
covariances of local probe signals over repeated preparations of randomised initial states, using forward evolution and probe readout only. In a spin-1 chain, a nonzero imaginary signal distinguishes a real Hamiltonian $H$ from $-H$ despite identical corresponding OTOCs, and identifies opposite sides of a pair-conversion resonance at short times while dipolar and quadrupolar OTOCs remain nearly unchanged.
These results show how temporal interference resolves dynamical ambiguities and provides additional constraints for Hamiltonian inference.

\end{abstract}

\maketitle

Quantum information spreads through interacting systems as initially local information becomes encoded in many-body correlations.
Characterising this process is a central task in nonequilibrium physics and quantum simulation.
Out-of-time-order correlators (OTOCs) provide a widely used diagnostic by tracking the growing noncommutativity of initially separated
observables
\cite{LarkinOvchinnikov1969,ShenkerStanford2014,Maldacena2016}, and have been measured across several experimental platforms
\cite{Garttner2017,Li2017,Landsman2019,Joshi2020,Blok2021,Mi2021,GoogleQuantumAI2025}. However, individual correlators retain only selected combinations
of the underlying outcome statistics.  
A richer statistical description can preserve interference information lost in these averages
\cite{Lostaglio2023,ArvidssonShukur2024},
potentially distinguishing quantum evolutions that conventional correlators leave unresolved. 

The extended Kirkwood-Dirac (KD) quasiprobability $Q$ provides such a description by resolving the outcome contributions underlying
an OTOC, as shown by Yunger Halpern and coworkers~\cite{YungerHalpern2017,YungerHalpern2018}.
Its negative and nonreal values reveal departures from classical
joint probabilities, with established connections to scrambling
witnesses and entropic uncertainty relations
\cite{GonzalezAlonso2019,YungerHalpern2019,thio2026kirkwood,budiyono2023general}.
At infinite temperature, standard OTOCs formed from two repeated
Hermitian observables depend only on weighted sums of
$\operatorname{Re}Q$; the imaginary component cancels.
This leaves a component of the underlying quantum interference
unresolved by these diagnostics.
These results motivate three questions: which physical distinctions are encoded in
$\operatorname{Im}Q$ but absent from conventional OTOC, when can $Q$ have a nonzero imaginary part at infinite temperature, and how can $Q$ be measured without reversing or measuring the system?

\begin{figure}[tbp]
\includegraphics[width=0.95\columnwidth]{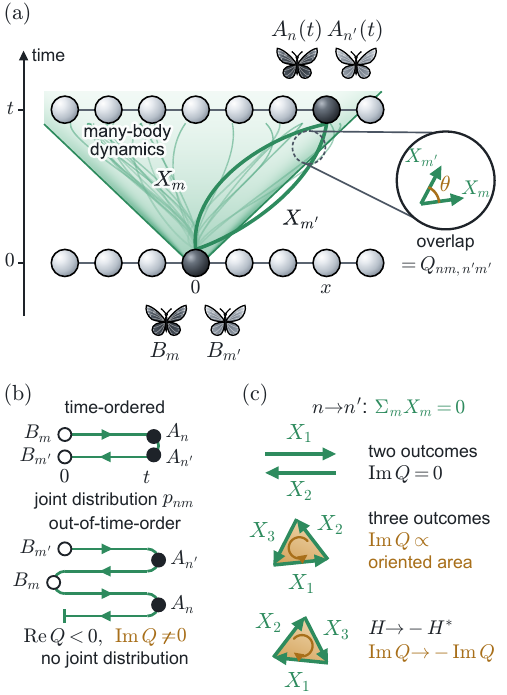}
\caption{Echo interpretation of $Q$. (a) The echo amplitudes $X_m :=A_{n'}(t)B_mA_n(t)$ for a transition $n\to n'$, with overlap $Q_{nm,n'm'}$. The light cone illustrates operator spreading. (b) Operator ordering for the classical marginal $p_{nm}$ and the extended KD quasiprobability $Q$. (c) The amplitudes for two and three outcomes. For three outcomes, the oriented triangle area defined in
SM Sec. \ref{S:prop1} is proportional to $\operatorname{Im}Q$; $H\mapsto-H^*$ reverses its sign.}
\label{fig:concept}
\end{figure}

Here we find that temporal correlations between measurements at different times can contain interference that no joint probability distribution reproduces. To that end, we show that $Q$ carries information about quantum dynamics that standard Hermitian OTOCs discard, and its entries are overlaps between echo amplitudes [Fig.~\ref{fig:concept}(a)]. In an undisturbed echo, amplitudes leading to a different final outcome cancel. Negative real parts of $Q$ describe the destructive interference that enforces this cancellation, while its imaginary part gives the complementary interference quadrature. We prove that this imaginary part vanishes for every unitary dynamics if either measurement has at most two outcomes, but is nonzero for generic unitary dynamics when both have at least three. Earlier quasiprobability schemes for measuring these overlaps use time reversal or ancillary control of the many-body evolution~\cite{YungerHalpern2017,YungerHalpern2018,Dressel2018,Mohseninia2019,Wagner2024}. Randomized OTOC protocols avoid time reversal but still need direct measurements on the system~\cite{Vermersch2019,Joshi2020}. By contrast, our protocol uses a coherently controlled local qubit to obtain both responses and symmetrized correlations~\cite{Dong2026}. We estimate these signals on repeated copies of each random state and take their covariances across states. This reconstructs $Q$ while the system evolves only forward in time and is never measured. A spin-1 chain shows why the additional information matters. Its dipolar and quadrupolar OTOCs can remain nearly unchanged while the imaginary part of $Q$ distinguishes opposite sides of a pair-conversion resonance at short times. A nonzero imaginary part also separates $H$ and $-H$ for this real Hamiltonian, although their Hermitian OTOCs coincide. The phase thus supplies a spectroscopic constraint on dynamics that standard OTOC data leave unresolved.

\textit{Classical and nonclassical correlations in time.}---We work at infinite temperature, and write $\Tr X\equiv\tr X/\mathcal D$ for the trace normalized by the Hilbert-space dimension $\mathcal D$. A projective measurement of a local observable at site $0$ and time $0$ has $s$ outcomes $m=1,\dots,s$ and projectors $B_m=P^{(m)}_0$. A second measurement at site $x$ and time $t$ has $r$ outcomes $n=1,\dots,r$ and Heisenberg projectors $A_n(t)=U^\dagger P^{(n)}_xU$, with $U=e^{-\ii Ht}$. The letters recall the order of the two measurements, $B$ before and $A$ after; the site $x$ is fixed throughout.

The extended KD quasiprobability of the two measurements is
$Q_{nm,n'm'}=\Tr(A_n(t)B_mA_{n'}(t)B_{m'})$. For observables $a=\sum_na^{(n)}P^{(n)}$ and $b=\sum_mb^{(m)}P^{(m)}$, the OTOC expands as~\cite{YungerHalpern2018} 
\begin{equation}
\begin{aligned}
F&=\Tr[a_x(t)b_0a_x(t)b_0]\\
&=\sum_{nmn'm'}a^{(n)}b^{(m)}a^{(n')}b^{(m')}\,Q_{nm,n'm'}.
\end{aligned}
\label{eq:Q}
\end{equation} 
Its entries sum to one and summing over $n'$ and $m'$ gives $p_{nm}=\Tr(A_n(t)B_m)\geq0$, i.e., the probability of obtaining $m$ and then $n$ when the two measurements are performed in sequence. Thus $Q$ has the marginals of a joint distribution of the outcomes, and the OTOC is one weighted sum of its entries. Time-ordered correlations of the same two measurements reduce to $p$. Arranging the four projectors in time order, for example, gives $\Tr(B_{m'}A_{n'}(t)A_n(t)B_m)=\delta_{nn'}\delta_{mm'}\,p_{nm}$, because the projectors of one measurement are mutually orthogonal and the trace is cyclic. The same reduction applies to time-ordered correlations of any order. All of them are moments of a single pair of classical outcomes distributed according to $p$ (SM Sec.~\ref{S:prop1}), so they carry no information beyond a joint distribution [Fig.~\ref{fig:concept}(b)].

The out-of-time-order product in Eq.~\eqref{eq:Q} admits no such reduction, because no two projectors of the same measurement become adjacent under cyclic shifts. The marginals of $Q$ constrain where a probability distribution could place its weight. Summed over the two late (early) outcomes, $Q$ vanishes unless $m=m'$ ($n=n'$) (SM Sec. \ref{S:conv}). A nonnegative distribution with these marginals is therefore confined to entries with $n=n'$ and $m=m'$. Completeness relates the entries in which both labels change to the incompatibility of the measurements
\begin{equation}
\sum_{n\neq n',\,m\neq m'}\operatorname{Re}Q_{nm,n'm'}=-\frac12\sum_{nm}\big\|[A_n(t),B_m]\big\|^2,
\label{eq:sumrule}
\end{equation} with $\|X\|^2=\Tr(X^\dagger X)$. Up to its sign, the r.h.s. is the squared mutual eigenspace disturbance~\cite{Gao2023}. Thus incompatible measurements force at least one entry with $n\neq n'$ and $m\neq m'$ to have $\operatorname{Re}Q_{nm,n'm'}<0$, and hence $Q$ is not a probability distribution [Fig.~\ref{fig:concept}(b)].

Besides negative entries, $Q$ can have complex ones. Hermitian conjugation reverses the order of the four projectors, and after a cyclic shift this is equivalent to exchanging the two outcomes of either measurement $Q^*_{nm,n'm'}=Q_{nm',n'm}=Q_{n'm,nm'}$. $\operatorname{Re}Q$ is thereby symmetric under these exchanges, whereas $\operatorname{Im}Q$ is antisymmetric. Explicitly, $2\ii\operatorname{Im}Q_{nm,n'm'}=\Tr(A_n(t)B_mA_{n'}(t)B_{m'})-\Tr(B_{m'}A_{n'}(t)B_mA_n(t))$ is the difference between the product of the four projectors and the same product in reverse order, and it vanishes whenever the projectors commute. This mirrors the ordinary KD quasiprobability $\tr(\rho A_n(t)B_m)$, whose imaginary part is the expectation of a commutator. Yet at infinite temperature the ordinary quasiprobability reduces to $p_{nm}$, so all nonclassical values reside in $Q$. The weights in Eq.~\eqref{eq:Q} are symmetric under the exchange, so the standard OTOC of Hermitian observables retains only $\operatorname{Re}Q$. The imaginary part requires either operators whose eigenvalues do not lie on a line, such as clock operators, or four independently chosen operators (SM Sec. \ref{S:ImQ}). We next explain the interference represented by $Q$ and the conditions under which its imaginary part survives.

\textit{Phase between echo amplitudes.}---To see where the nonclassical entries of $Q$ come from, fix two different outcomes $n\neq n'$ of the late measurement and define $X_m :=A_{n'}(t)B_mA_n(t)$. As in an OTOC, $X_m$ describes an echo: forward to time $t$, back to $0$ and forward again, through the outcomes $n$, $m$ and $n'$ in turn [Fig.~\ref{fig:concept}(b)]. Since $n'\neq n$, $X_m$ is the amplitude for the echo to change the late outcome; they describe alternative paths through the early outcomes. By definition, $Q_{nm,n'm'}=\Tr(X_m^\dagger X_{m'})$ is their overlap [Fig.~\ref{fig:concept}(a)]. Combining two paths coherently gives the transition weight $\|X_m+e^{\ii\phi}X_{m'}\|^2$, which contains the interference term $2\operatorname{Re}(e^{\ii\phi}Q_{nm,n'm'})$. The real and imaginary parts of $Q$ are hence the two quadratures of this interference. %
Completeness gives $\sum_mX_m=A_{n'}(t)A_n(t)=0$: without the early measurement, the forward and backward evolutions cancel, and the echo cannot change the late outcome. Performing the early measurement and discarding its outcome instead gives the nonnegative transition weight $\sum_m\|X_m\|^2$ from $n$ to $n'$. The interference between different early outcomes cancels this weight $\sum_{m\neq m'}\operatorname{Re}Q_{nm,n'm'}=-\sum_m\|X_m\|^2$, so the entries of $Q$ with negative real parts are the destructive interference required for an undisturbed echo to return to its initial outcome. Summing over $n\neq n'$ recovers Eq.~\eqref{eq:sumrule}.

For an early measurement with only two outcomes, cancellation forces $X_2=-X_1$, so their overlap is real and nonpositive. With three outcomes, however, the amplitudes can close into a triangle with complex overlaps [Fig.~\ref{fig:concept}(c)], whose imaginary parts are equal for cyclically ordered pairs of sides and proportional to the oriented area of the triangle. The same argument applies with the two measurements exchanged. The above nonclassical properties are summarized below (SM Sec. \ref{S:prop1} and \ref{S:theorem} for detailed proof).

\begin{proposition}[Non-classicality of $Q$] At infinite temperature: (i) Every time-ordered correlation of the two measurements is a moment of the joint distribution $p_{nm}$. (ii) $Q$ is real and nonnegative iff $[A_n(t),B_m]=0$ $\forall n,m$. Otherwise, Eq. \eqref{eq:sumrule} requires entries with negative real parts. (iii) For $n\ne n'$, $Q_{nm,n'm'}=\Tr(X_m^\dagger X_{m'})$. Its real and imaginary parts are the two quadratures of interference between the echo paths. Entries with $n=n'$ or $m=m'$ are real.
\label{prop:prop1}
\end{proposition}

Combined with the cancellation argument above, (iii) gives our first main result.

\begin{theorem}[Three-outcome theorem, informal] At infinite temperature, $Q$ is real for every unitary dynamics if one of the two measurements has at most two outcomes. If both measurements have at least three outcomes, $Q$ is complex for generic unitary dynamics.
\label{thm:Thm1}
\end{theorem}

The projectors may have any rank, so Theorem \ref{thm:Thm1} applies to local measurements in a many-body system. For binary measurements, $Q$ is real irrespective of the dynamics, although incompatible measurements still produce negative entries. This establishes, for any binary measurement, the real $Q$ observed numerically for Pauli operators at infinite temperature and attributed there to a cancellation~\cite{YungerHalpern2018}. A spin-1 is thereby the smallest local system whose $Q$ can have a nonzero imaginary part. What counts is the number of resolved outcomes: two qubits read out as one four-level site also qualify. The orientation of the triangle has a simple dynamical meaning. Replacing $H$ by $-H^*$, with complex conjugation taken in the basis of the measured levels, maps $Q$ to $Q^*$ (SM Sec. \ref{S:reversal}): the triangle reverses its orientation and $\operatorname{Im}Q$ changes sign, while $\operatorname{Re}Q$ is unchanged [Fig.~\ref{fig:concept}(c)]. For a Hamiltonian that is real in this basis, $-H^*=-H$ generates the same dynamics run backwards. Whenever $\operatorname{Im}Q\neq0$, its sign change distinguishes $H$ from $-H^*$, while the real part and the marginals $p_{nm}$ remain unchanged.

\begin{figure}[tbp]
\includegraphics[width=\columnwidth]{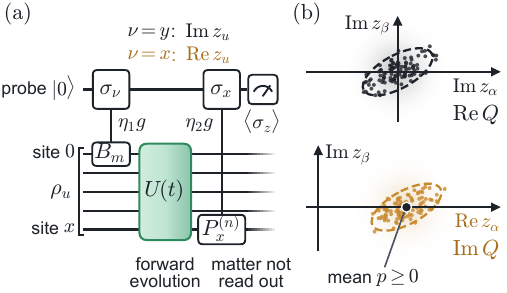}
\caption{Probe reconstruction of $Q$.
(a) Couplings to $B_m$ and $P_x^{(n)}$ separated by forward evolution $U(t)$, with signed pulse areas $\eta_1g$ and $\eta_2g$. Each state $\rho_u$ is drawn from a global projective $2$-design and prepared again for every shot and setting; only the probe
is read out. (b) Schematic: covariances across states. Together with $p_{nm}$, response-response covariances determine
$\operatorname{Re}Q$; mixed-quadrature covariances determine $\operatorname{Im}Q$.
The clouds are centered at $(0,0)$ above and $(p_\alpha,0)$ below.}
\label{fig:protocol}
\end{figure}

\textit{Learning $Q$ with a local probe.}---We reconstruct $Q$ by converting a trace of four projectors into correlations between two-time signals. For each randomly prepared pure state $\rho_u$, define $z_{u,nm}=\langle A_n(t)B_m\rangle_u$. Averaging over a global projective 2-design~\cite{Dankert2009} gives $p_{nm}=\E_uz_{u,nm}$ and 
\begin{equation}
Q_{nm,n'm'}=p_{nm}p_{n'm'}+(\mathcal D+1)\Cov_u\big(z_{u,nm},z_{u,n'm'}\big),
\label{eq:randomstates}
\end{equation} 
where the covariance contains no complex conjugation (SM Sec.~\ref{S:probe}). For each label $u$, independent copies of the same state must be prepared to estimate all $z_{u,nm}$; their covariance is then taken across state labels [Fig.~\ref{fig:protocol}]. Thus randomization supplies the trace identity, while repeated preparation preserves the correlations needed to use it. The factors $z_{u,nm}$ are two-time KD quasiprobabilities of the prepared states. Their imaginary part
$\operatorname{Im}z_{u,nm}
=-\tfrac12\langle\ii[A_n(t),B_m]\rangle_u$
is a response, whereas their real part
$\operatorname{Re}z_{u,nm}
=\tfrac12\langle\{A_n(t),B_m\}\rangle_u$
is a symmetrized correlation. Together with $p_{nm}$, response covariances determine
$\operatorname{Re}Q$; reconstructing $\operatorname{Im}Q$ also requires covariances between responses and symmetrized correlations [Fig.~\ref{fig:protocol}(b)]. Our quantum probe supplies both quadratures without reading out the many-body system \cite{Dong2026}.

In our algorithm [Fig.~\ref{fig:protocol}(a); details in SM Sec. \ref{S:probe}], the probe starts in $\ket{0}$ and couples first to $B_m$ at site $0$ via $\exp(-\ii\eta_1g\sigma_\nu\otimes B_m)$. After forward evolution for time $t$, it couples to $P^{(n)}_x$ via $\exp(-\ii\eta_2g\sigma_x\otimes P^{(n)}_x)$ and is measured through $\sigma_z$. Here $g$ is a weak coupling strength, and $\eta_1,\eta_2\in\{-1,+1\}$ are independently chosen signs for the two pulses. The late coupling uses the probe axis $\sigma_x$: choosing the early axis $\sigma_\nu$ with $\nu=y$ or $x$ selects the response or the symmetrized correlation. For each axis, four choices of the two pulse signs are measured in separate runs. Their sign-weighted combination $s_\nu=\frac14\sum_{\eta_1\eta_2}\eta_1\eta_2\langle\sigma_z\rangle_{\eta_1\eta_2}$ isolates the contribution involving both couplings and gives 
\begin{equation}
\begin{aligned}
s_y &= 4g^2\operatorname{Im}z_{u,nm}+O(g^4),\\
s_x &= -4g^2\operatorname{Re}z_{u,nm}+O(g^4).
\end{aligned}
\end{equation}
These signals determine the means and covariances in Eq.~\eqref{eq:randomstates}. Measuring both axes for all outcome pairs requires $8rs$ settings per random state. The required resources are random-state preparation, calibrated local probe couplings, and probe readout. The many-body system undergoes only forward evolution and needs no readout, which suits systems that cannot be measured site by site, such as a nuclear-spin register addressed through an electron spin~\cite{Randall2021,Meinel2022}. Finite coupling introduces an $O(g^2)$ bias, while sampling error decreases as $R^{-1/2}$ with the number $R$ of random states. For Haar-random states, bounded operators and target signals that
do not shrink with system size, the required number of random states need not grow with $\mathcal D$.
Resolving each record nevertheless requires a shot budget $N_s\propto \mathcal D/g^2$ per setting.
The resulting preparation cost and error budget are given in SM Sec. \ref{S:errors}.

\begin{figure}[tbp]
\includegraphics[width=\columnwidth]{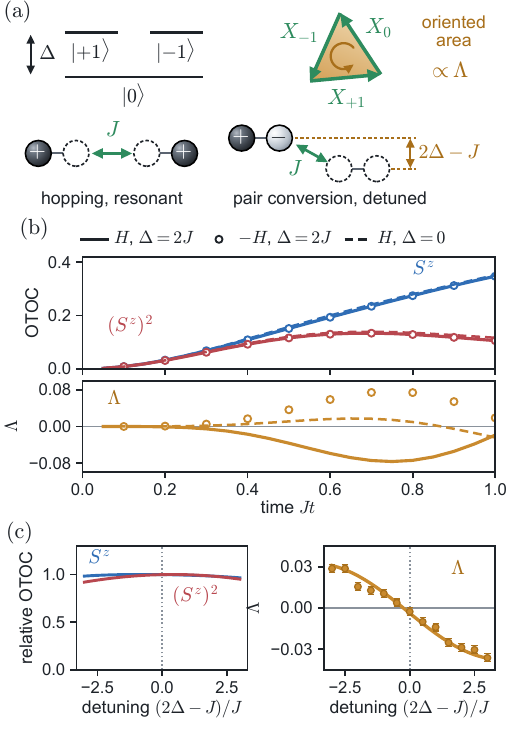}
\caption{Spin-1 dynamics and probe reconstruction. (a) Local levels, echo amplitudes and bond processes. (b) Squared-commutator OTOCs
$\|[a_x(t),a_0]\|^2$ for $a=S^z$ and $a=(S^z)^2$, and $\Lambda$. (c) OTOCs normalized by their values at resonance and $\Lambda$, at $Jt=0.5$. All curves use $x=1$ and $L=6$. Points are probe estimates of $\Lambda$ with statistical
standard errors, each using $L=5$, $g=0.15$,
$2\times10^5$ random states, and $10^4$ shots per sign setting.}
\label{fig:spin1}
\end{figure}

\textit{Paths and phases in a spin-1 chain.}---For a spin-1, we take both measurements to resolve the three levels of $S^z$. These outcomes resolve a dipole, $S^z$, which distinguishes
$\ket{+1}$ from $\ket{-1}$, and a longitudinal quadrupole, proportional to $(S^z)^2-\tfrac23 \mathbb{I}$, which distinguishes the magnetic levels from $\ket{0}$. $\operatorname{Im}Q$ then has a single coordinate, which we define by
$\Lambda=-54 \operatorname{Im}Q_{+1,+1;\,0,0}$.
Every entry in which both outcomes change has
$\operatorname{Im}Q=\pm\Lambda/54$, with the sign determined by the orientations of the two transitions in the cyclic order
$+1\to0\to-1\to+1$ (SM Sec. \ref{S:spin1}). The coordinate $\Lambda$ is thus proportional to the oriented area of the triangle of echo amplitudes [Fig.~\ref{fig:spin1}(a)]. We consider the spin-1 chain~\cite{Chung2021} \begin{equation}
H=J\sum_j\Big[S^z_jS^z_{j+1}+\tfrac12\big(S^+_jS^-_{j+1}+\mathrm{h.c.}\big)\Big]+\Delta\sum_j(S^z_j)^2.
\label{eq:H}
\end{equation} 
The levels $\ket{\pm1}$ can be viewed as particles with two flavours, and $|0\rangle$ as a vacancy. The exchange term moves a particle onto a neighboring vacancy without changing the bond energy and converts a pair $|{+}1,{-}1\rangle$ into $|0,0\rangle$, a process detuned by $2\Delta-J$ [Fig.~\ref{fig:spin1}(a)]. The anisotropy $\Delta$ thus tunes the balance between the two processes. Fig.~\ref{fig:spin1}(b) shows that the OTOCs barely register this balance. On neighboring sites and for $Jt\leq0.8$, the squared-commutator OTOCs of $S^z$ and $(S^z)^2$ at $\Delta=2J$ and at $\Delta=0$ are almost identical, and reversing the dynamics $H\to-H$ leaves them unchanged. In contrast, $\Lambda$ reverses under $H\to-H$, as anticipated in Fig.~\ref{fig:concept}(c), and it has opposite signs at $\Delta=2J$ and at $\Delta=0$. Dynamics with nearly equal OTOCs thus differ in $Q$ through its imaginary part.

The imaginary coordinate probes the phase accumulated along conversion paths. On a bipartite chain, $\Lambda$ vanishes when the diagonal $S_j^zS_{j+1}^z$ and $(S_j^z)^2$ terms are absent (SM Sec. \ref{S:spin1}). For neighboring sites, $\Lambda=J^4(J-2\Delta)t^5+O(t^7)$: its leading coefficient measures the bond detuning $2\Delta-J$ with the opposite sign, and hence distinguishes which side of the conversion resonance the dynamics occupies. Across the resonance $\Delta=J/2$, the OTOCs are approximately even in the detuning, whereas $\Lambda$ is approximately odd [Fig.~\ref{fig:spin1}(c)]. At finite time, higher-order contributions shift its zero crossing, so the sign of $\Lambda$ is a detuning diagnostic in the short-time regime. We test the reconstruction using the calibrated linear and quadratic
spin couplings specified in SM Sec. \ref{S:spin1}.
Two response records obtained from independent shots give the OTOC of $S^z$, while pairing one response record with a symmetrized-correlation
record gives $\Lambda$. Three records per random state hence recover both quantities
without reconstructing every entry of $Q$.
Fig.~\ref{fig:spin1}(c) shows the reconstructed $\Lambda$ across the conversion resonance.

\textit{Discussion and outlook.}---Our results connect the negative and nonreal entries of quasiprobability
$Q$ to interference between many-body echo paths and make both quadratures accessible through a local probe. The classification gives experiments a concrete design rule for revealing dynamical information beyond time correlators. At infinite temperature, binary measurements can reveal negative weights, but imaginary echo overlaps require at least three resolved outcomes in each measurement. A real $Q$ for local qubit measurements therefore does not imply classical dynamics. Three-level sites and jointly resolved qubit registers provide access to imaginary overlaps excluded by binary measurements. This information distinguishes Hamiltonians that are indistinguishable based on conventional data. For the fixed-basis observables considered here, $H$ and $-H^*$ produce identical time-ordered correlations and Hermitian OTOCs at infinite temperature, whereas $\operatorname{Im}Q$ changes sign. A nonzero imaginary signal hence distinguishes these candidates, removing a sign ambiguity for real Hamiltonians and adding a constraint to learning from that data~\cite{Bairey2019,Anshu2021,Huang2023,GoogleQuantumAI2025}. The spin-1 example gives this constraint a concrete spectroscopic meaning: the short-time signal identifies the side of the conversion resonance, in contrast to almost unchanged dipolar and quadrupolar OTOCs.

Experimentally, covariances of forward probe signals recover echo interference without reversing the system dynamics. This is useful when the system can be repeatedly prepared and evolved, but readout is restricted to an ancillary system. Repeatable random-state preparation and calibrated probe couplings remain essential. Qutrit processors~\cite{Blok2021}, spin-1 atoms with tunable anisotropy~\cite{Chung2021}, and nuclear-spin registers addressed through an electron spin~\cite{Randall2021,Meinel2022} offer candidate platforms for its probe, which reveal the interference beyond OTOC. Reducing the cost of global random-state preparation and probe sampling is a priority for extending these measurements to larger systems. Measurements with more outcomes provide further imaginary coordinates, opening the possibility of resolving several independent interference signals. Extending the protocol to thermal states is another open problem: initial-state coherences in the measured basis can make $Q$ nonreal even for binary measurements, and their contribution must be accounted for when interpreting the measured phase.

\textit{Acknowledgments.---}W.D. thanks Haoyu Guan for discussions. A.Y. is supported by UKRI Future Leaders Fellowship (Grant No. 10128920).

\bibliography{references}

\clearpage
\onecolumngrid
\begin{center}
{\large\textbf{Supplemental Materials for\\ [0.30em]
"Complex quasiprobabilities as measurable temporal interference"}}
\end{center}
\vspace{2mm}
Sec.~\ref{S:conv}-\ref{S:reversal} hold for any unitary dynamics, Sec.~\ref{S:probe}-\ref{S:errors} concern the probe algorithm and the spin-1 chain, and Sec.~\ref{S:finiteT} states which results hold beyond infinite temperature.

\section{Conventions and marginals}
\label{S:conv}

Each site carries a complete set of orthogonal level projectors $\sum_nP^{(n)}=\id$ and $P^{(n)}P^{(n')}=\delta_{nn'}P^{(n)}$. The projectors $A_n$ ($n=1,\dots,r$) and $B_m$ ($m=1,\dots,s$) are therefore complete orthogonal families on the full Hilbert space (of arbitrary rank), and only nonzero projectors are counted. Traces are normalized $\Tr\id=1$ and $\|X\|^2=\Tr(X^\dagger X)$. We write $A_n$ for the late projectors $A_n(t)$ of the main text. Until Sec.~\ref{S:probe}, the results use neither the form of $H$ nor locality and hold for any unitary $U$. Completeness of the two families gives the one-label sums
\begin{equation}
\begin{aligned}
\sum_nQ_{nm,n'm'}&=\delta_{mm'}\,p_{n'm},&\qquad\sum_{n'}Q_{nm,n'm'}&=\delta_{mm'}\,p_{nm},\\
\sum_mQ_{nm,n'm'}&=\delta_{nn'}\,p_{nm'},&\qquad\sum_{m'}Q_{nm,n'm'}&=\delta_{nn'}\,p_{nm},
\end{aligned}
\label{eq:S-marginals}
\end{equation}
where each three-projector trace has been reduced by cyclicity and orthogonality. For the first, $\sum_nQ_{nm,n'm'}=\Tr(B_mA_{n'}B_{m'})=\Tr(B_{m'}B_mA_{n'})=\delta_{mm'}\Tr(B_mA_{n'})$. The other three follow in the same way. Summing once more gives $\sum_{nn'}Q_{nm,n'm'}=\delta_{mm'}\Tr B_m$ and $\sum_{mm'}Q_{nm,n'm'}=\delta_{nn'}\Tr A_n$. A nonnegative distribution with these marginals must hence vanish whenever $m\neq m'$ or $n\neq n'$, because a sum of nonnegative terms vanishes only if every term does. Because the projectors are Hermitian, $Q^*_{nm,n'm'}=\Tr(B_{m'}A_{n'}B_mA_n)$, and the two cyclic shifts of this product give
\begin{equation}
Q^*_{nm,n'm'}=Q_{nm',n'm}=Q_{n'm,nm'}.
\label{eq:S-conj}
\end{equation}
$\operatorname{Re}Q$ is hence symmetric and $\operatorname{Im}Q$ antisymmetric under $m\leftrightarrow m'$ and under $n\leftrightarrow n'$. In particular, $\operatorname{Im}Q_{nm,n'm'}=0$ if $n=n'$ or $m=m'$, and by Eq.~\eqref{eq:S-marginals} every one-label sum of $\operatorname{Im}Q$ vanishes.

\section{Proof of Proposition 1}
\label{S:prop1}

Proposition \ref{prop:prop1} of the main text contains three statements, proved here as individual Lemmas: (i) time-ordered correlations reduce to one distribution, (ii) the out-of-time-order kernel has negative weight exactly when the two measurements are incompatible, and (iii) its imaginary part is the oriented area spanned by echo amplitudes. (i) and (ii) build on \cite{YungerHalpern2018,Gao2023}. Table~\ref{tab:classes} summarizes the resulting classification.

\begin{table}[!htb]
\centering
\caption{Correlations of two measurements at two times, at infinite temperature. TO: time-ordered; OTO: out-of-time-order. Compatible means $[A_n,B_m]=0$ $\forall n,m$, and $r$ and $s$ are the numbers of outcomes of the late and early measurements. The last row is Theorem \ref{thm:Thm1}; ``generically'' refers to Haar-random unitary dynamics.}
\label{tab:classes}
\begin{tabular}{@{}lccc@{}}
\toprule
Kernel & $\operatorname{Re}Q$ & $\operatorname{Im}Q$ & Joint distribution\\
\midrule
TO, any order & $\geq0$ & $0$ & yes\\
OTO, compatible & $\geq0$ & $0$ & yes\\
OTO, incompatible, $\min(r,s)=2$ & some $<0$ & $0$ & no\\
OTO, incompatible, $r,s\geq3$ & some $<0$ & generically $\neq0$ & no\\
\bottomrule
\end{tabular}
\end{table}

\begin{lemma}
\label{lem:TOC}
For any $k,l\geq1$ and any labels
\begin{equation}
\Tr\big(A_{n_1}\cdots A_{n_k}B_{m_1}\cdots B_{m_l}\big)=\Big(\prod_{j=2}^{k}\delta_{n_jn_1}\Big)\Big(\prod_{j=2}^{l}\delta_{m_jm_1}\Big)\,p_{n_1m_1}.
\label{eq:S-TOC}
\end{equation}
\end{lemma}
\begin{proof}
Adjacent projectors of the same family multiply as $A_nA_{n'}=\delta_{nn'}A_n$ and $B_mB_{m'}=\delta_{mm'}B_m$, which leaves $\Tr(A_{n_1}B_{m_1})=p_{n_1m_1}$.
\end{proof}
A correlator of insertions at the two times on a single forward-backward time contour has, up to a cyclic shift, one block of $A$'s and one of $B$'s, so Lemma~\ref{lem:TOC} applies. Let $V=\sum_mv_mB_m$ and $W=\sum_nw_nA_n$ be operators diagonal in the two measurements, and let $f$ and $g$ be functions. Every such correlator then reduces to
\begin{equation}
\Tr[f(W)\,g(V)]=\sum_{nm}f(w_n)\,g(v_m)\,p_{nm},
\label{eq:S-moments}
\end{equation}
where $f(W)$ collects all insertions at the late time and $g(V)$ all insertions at the early time. For example, $\Tr(VW^2V)=\sum_{nm}w_n^2v_m^2p_{nm}$. Correlators of this kind are therefore moments of a pair of classical random variables $(w,v)$ with joint distribution $p_{nm}$, whether the operators are Hermitian or not. For a state $\rho$ that commutes with every $B_m$, the time-ordered counterpart of $Q$ reduces in the same way, $\tr(\rho B_mA_nA_{n'}B_{m'})=\delta_{nn'}\delta_{mm'}\tr(\rho B_mA_n)$, and $\tr(\rho B_mA_n)$ is a probability~\cite{YungerHalpern2018}. Correlators with more than two times, or with noncommuting operators at the same time, are not covered by this statement. The out-of-time-order product in $Q$ places the two times alternately and is not of this form.

\begin{lemma}
\label{lem:neg}
Eq.~\eqref{eq:sumrule} holds. Consequently, $Q$ is real and nonnegative if and only if $[A_n,B_m]=0$ $\forall n,m$, and this is already implied by $\operatorname{Re}Q_{nm,n'm'}\geq0$ for all entries. If some $A_n$ does not commute with some $B_m$, an entry with $n\neq n'$ and $m\neq m'$ has a negative real part.
\end{lemma}
\begin{proof}
Let $S=\sum_{nm}Q_{nm,nm}$. By the one-label sums, the sum of all entries, the sum over entries with $n=n'$, and the sum over entries with $m=m'$ each equal one. By inclusion and exclusion, the entries with $n\neq n'$ and $m\neq m'$ sum to $S-1$. Expanding the commutator gives $\|[A_n,B_m]\|^2=2p_{nm}-2Q_{nm,nm}$ because $\Tr(B_mA_nB_m)=\Tr(A_nB_mA_n)=p_{nm}$ and cyclicity makes $\Tr(A_nB_mA_nB_m)$ and $\Tr(B_mA_nB_mA_n)$ equal. In particular, $Q_{nm,nm}$ is real. Summing over $n,m$ gives $2-2S$, which proves Eq.~\eqref{eq:sumrule}. If all commutators vanish, $Q_{nm,n'm'}=\Tr(A_nA_{n'}B_mB_{m'})=\delta_{nn'}\delta_{mm'}p_{nm}$. If all real parts are nonnegative, the left-hand side of Eq.~\eqref{eq:sumrule} is nonnegative while the right-hand side is not positive, so all commutators vanish.
\end{proof}
Entries in which exactly one label changes are real and nonnegative. For $m\neq m'$, $Q_{nm,nm'}=\Tr[(A_nB_mA_n)(A_nB_{m'}A_n)]$ is the trace of a product of two positive operators, and likewise for $n\neq n'$. Negative real parts therefore occur only where both labels change, and Eq.~\eqref{eq:sumrule} bounds the total negative weight $\sum\max(0,-\operatorname{Re}Q_{nm,n'm'})\geq\frac12\sum_{nm}\|[A_n,B_m]\|^2$. In the notation of Ref.~\cite{Gao2023}, $S=1-\mathrm{MED}^2$, where MED is the mutual eigenspace disturbance of the two measurements. Eq.~\eqref{eq:sumrule} thus restates their identity in terms of $Q$: the entries in which both labels change sum to $-\mathrm{MED}^2$. Individual entries in which both labels change need not be negative. In the spin-1 chain at $\Delta=2J$, $x=1$ and $Jt=0.7$ ($L=6$), for example, the four entries in which both transitions connect the levels $+1$ and $-1$ have real part $+3\times10^{-4}$, whereas the entries in which both labels change sum to $-0.32$.

For $n\neq n'$ let $X_m=A_{n'}B_mA_n$. Then $X_m^\dagger X_{m'}=A_nB_mA_{n'}B_{m'}A_n$, and cyclicity together with $A_n^2=A_n$ gives $Q_{nm,n'm'}=\Tr(X_m^\dagger X_{m'})$. For fixed $n\neq n'$ the entries $Q_{nm,n'm'}$ thus form the Gram matrix of the amplitudes $\{X_m\}$, which is positive semidefinite; its off-diagonal entries carry the signs and phases. Completeness of the early measurement gives $\sum_mX_m=0$, and expanding $\|\sum_mX_m\|^2=0$ gives $\sum_{m\neq m'}Q_{nm,n'm'}=-\sum_m\|X_m\|^2$. Its right-hand side, $\sum_m\|X_m\|^2=\Tr[A_{n'}\Phi_B(A_n)]$ with $\Phi_B(O)=\sum_mB_mOB_m$, is the weight with which a nonselective early measurement moves the late outcome from $n$ to $n'$; summed over $n\neq n'$ it equals $\mathrm{MED}^2$, consistent with Lemma~\ref{lem:neg}. Exchanging the roles of the two measurements, the amplitudes $Y_n=B_mA_nB_{m'}$ with $m\neq m'$ satisfy $\Tr(Y_n^\dagger Y_{n'})=Q_{nm,n'm'}$ and $\sum_nY_n=0$. Regard the operators as a real vector space and define on it the antisymmetric bilinear form $\omega(X,Y)=\operatorname{Im}\Tr(X^\dagger Y)$, the oriented area spanned by $X$ and $Y$. Then $\operatorname{Im}Q_{nm,n'm'}=\omega(X_m,X_{m'})$. To see why $\omega$ is an area, expand $X$ and $Y$ in an orthonormal operator basis $\{E_k\}$, $\Tr(E_k^\dagger E_l)=\delta_{kl}$, with complex components $x_k$ and $y_k$. Then $\omega(X,Y)=\sum_k\operatorname{Im}(x_k^*y_k)=\sum_k(\operatorname{Re}x_k\operatorname{Im}y_k-\operatorname{Im}x_k\operatorname{Re}y_k)$, and each term is the signed area of the parallelogram spanned by $x_k$ and $y_k$ in the complex plane. The form $\omega$ is thus the sum of the oriented areas of the projections of the parallelogram spanned by $X$ and $Y$ onto the complex coordinate planes, the symplectic area familiar from Hamiltonian mechanics. Its modulus is bounded by the Euclidean area, $|\omega(X,Y)|\leq\|X\|\|Y\|\sin\phi$, where $\cos\phi=\operatorname{Re}\Tr(X^\dagger Y)/(\|X\|\|Y\|)$, with equality when $Y$ is a complex multiple of $X$; this follows from $|\Tr(X^\dagger Y)|\leq\|X\|\|Y\|$.
\begin{lemma}
\label{lem:phase}
$\operatorname{Im}Q\neq0$ if and only if $\omega(X_m,X_{m'})\neq0$ for some $n\neq n'$ and $m\neq m'$. Equivalently, $Q$ is real if and only if, for every transition $n\to n'$, the real span of the amplitudes $\{X_m\}$ is isotropic for $\omega$.
\end{lemma}
\begin{proof}
Entries with $n=n'$ or $m=m'$ are real by Eq.~\eqref{eq:S-conj}; the remaining entries are $\omega(X_m,X_{m'})$.
\end{proof}
For three early outcomes, closure gives $X_3=-X_1-X_2$, and bilinearity yields $\omega(X_1,X_2)=\omega(X_2,X_3)=\omega(X_3,X_1)=2\mathcal A$, where $\mathcal A$ is the oriented area, measured by $\omega$, of the triangle with sides $X_1$, $X_2$ and $X_3$. The three imaginary entries for the transition $n\to n'$ are thus all equal to $2\mathcal A$, and reversing the orientation of the triangle reverses their sign. When all projectors have rank one, $A_n=|a_n\rangle\langle a_n|$ and $B_m=|b_m\rangle\langle b_m|$, each entry
\begin{equation}
Q_{nm,n'm'}=\mathcal D^{-1}\langle a_n|b_m\rangle\langle b_m|a_{n'}\rangle\langle a_{n'}|b_{m'}\rangle\langle b_{m'}|a_n\rangle
\end{equation}
is a fourth-order Bargmann invariant, whose phase does not depend on the phases of the individual states~\cite{Wagner2024}. It is nonreal if and only if the four overlaps are nonzero and the phase of their product is not a multiple of $\pi$. For projectors of higher rank, $Q$ sums such products over the degenerate subspaces, and their phases can cancel. For example, on the doubled space $\mathcal H\oplus\mathcal H$ the families $A_n\oplus A_n^*$ and $B_m\oplus B_m^*$ have the same outcome numbers as $\{A_n\}$ and $\{B_m\}$, and their quasiprobability is $\frac12(Q+Q^*)=\operatorname{Re}Q$. Lemma~\ref{lem:phase} therefore has to be applied to the full trace rather than to a single product. Lemmas~\ref{lem:TOC}, \ref{lem:neg} and~\ref{lem:phase} are items (i), (ii) and (iii) of Proposition \ref{prop:prop1} in the main text.

\section{The three-outcome theorem}
\label{S:theorem}

\noindent\textbf{Theorem 1} (Three-outcome theorem, formal)\textbf{.} \textit{Let $\{A_n\}_{n=1}^r$ and $\{B_m\}_{m=1}^s$ be complete families of nonzero orthogonal projectors on a Hilbert space of dimension $\mathcal D$, and let $Q_{nm,n'm'}=\Tr(A_nB_mA_{n'}B_{m'})$.
(i) If $r\leq2$ or $s\leq2$, $Q$ is real.
(ii) Fix the family $\{B_m\}$ and the ranks of the $A_n$, and let $A_n=U^\dagger A_n^{(0)}U$ with $U$ ranging over the unitary group. If $r\geq3$ and $s\geq3$, then $\operatorname{Im}Q\neq0$ for all $U$ outside a set of Haar measure zero.}

\begin{proof}
(i) Let $s=2$. For $n\neq n'$, closure gives $X_2=-X_1$, so $Q_{n1,n'2}=Q_{n2,n'1}=-\|X_1\|^2$ is real. Entries with $n=n'$ or $m=m'$ are real by Eq.~\eqref{eq:S-conj}. The case $r=2$ follows in the same way from the amplitudes $Y_n$, and a family with a single outcome commutes with everything. (ii) We first construct one $U$ with $\operatorname{Im}Q\neq0$. Choose unit vectors $e_0$, $e_1$, $e_2$ in the ranges of $B_1$, $B_2$, $B_3$, respectively, and set $|a_j\rangle=3^{-1/2}\sum_{l=0}^{2}\omega_3^{lj}e_l$ for $j=0,1,2$, with $\omega_3=e^{2\pi\ii/3}$. Place $|a_j\rangle$ in $A_{j+1}$, and complete all $A_n$ to their prescribed ranks with eigenvectors of the $B_m$ in the orthogonal complement of $\mathrm{span}\{e_0,e_1,e_2\}$. This is possible because the required total dimension is exactly $\mathcal D-3$. On the complement the two families commute, so for $n\neq n'$ and $m\neq m'$ only the three-dimensional block contributes, and $Q_{12,21}=\mathcal D^{-1}\langle a_0|e_1\rangle\langle e_1|a_1\rangle\langle a_1|e_0\rangle\langle e_0|a_0\rangle=\omega_3/(9\mathcal D)$, whose imaginary part $\sqrt3/(18\mathcal D)$ is nonzero. The family so constructed has the prescribed ranks and is therefore of the form $U^\dagger A_n^{(0)}U$. The function $U\mapsto\operatorname{Im}Q_{12,21}$ is a polynomial in the entries of $U$ and $U^*$, hence real analytic on the connected group $\mathrm{U}(\mathcal D)$, and it is not identically zero. Its zero set therefore has Haar measure zero, and so has the set on which all of $\operatorname{Im}Q$ vanishes.
\end{proof}

For Pauli observables, the reality of $Q$ at infinite temperature was noted in \cite{YungerHalpern2018}, and it also follows from the expansion of the qubit quasiprobability into eight real correlators in \cite{Mohseninia2019}. Part~(i) extends it to any two-outcome measurement of any rank. Part~(ii) concerns the full unitary orbit. A particular Hamiltonian can still give a real $Q$, for instance when it commutes with the measured observables. The outcome numbers also bound the number of independent imaginary coordinates. By Eq.~\eqref{eq:S-conj} and the vanishing one-label sums, $\operatorname{Im}Q$ is antisymmetric with zero row sums in each pair of labels, which leaves at most $\binom{r-1}{2}\binom{s-1}{2}$ independent real coordinates. For $r=s=3$ there is a single one.

\section{Correlators that contain the imaginary part}
\label{S:ImQ}

For operators $V=\sum_mv_mP^{(m)}_0$ and $W=\sum_nw_nP^{(n)}_x$ with complex eigenvalues, the OTOC expands as
\begin{equation}
F(V,W)=\Tr[W^\dagger(t)V^\dagger W(t)V]=\sum_{nmn'm'}w_n^*v_m^*w_{n'}v_{m'}\,Q_{nm,n'm'}.
\end{equation}
The weights $c_{mm'}=v_m^*v_{m'}$ and $d_{nn'}=w_n^*w_{n'}$ are Hermitian matrices, so their real parts are symmetric and their imaginary parts antisymmetric. Combined with Eq.~\eqref{eq:S-conj}, only the terms of matching symmetry survive the sum
\begin{equation}
\operatorname{Re}F=\sum_{nmn'm'}\operatorname{Re}c_{mm'}\operatorname{Re}d_{nn'}\operatorname{Re}Q_{nm,n'm'},\qquad
\operatorname{Im}F=-\sum_{nmn'm'}\operatorname{Im}c_{mm'}\operatorname{Im}d_{nn'}\operatorname{Im}Q_{nm,n'm'}.
\label{eq:S-ReImF}
\end{equation}
For Hermitian $V$ and $W$ the weights are real and $F$ contains only $\operatorname{Re}Q$, as does the squared commutator $\Tr\{[W(t),V]^\dagger[W(t),V]\}=2\Tr[W^\dagger(t)W(t)VV^\dagger]-2\operatorname{Re}F$, whose first term is a time-ordered correlator. More generally, if the eigenvalues of $V$ lie on a line, $v_m=\alpha+\beta h_m$ with real $h_m$, then $\operatorname{Im}c_{mm'}=\operatorname{Im}(\alpha^*\beta)(h_{m'}-h_m)$, and its contraction with $\operatorname{Im}Q$ vanishes because the one-label sums of $\operatorname{Im}Q$ vanish. The same holds for $W$. Hence $\operatorname{Im}Q$ enters an OTOC only if neither the eigenvalues of $V$ nor those of $W$ lie on a common line, as for the clock operators with eigenvalues $e^{2\pi\ii k/3}$ used in qutrit experiments~\cite{Blok2021}. For the spin-1 chain, let $Z=\sum_j\omega_3^jP^{(m_j)}$ with $\omega_3=e^{2\pi\ii/3}$ and $(m_0,m_1,m_2)=(+1,0,-1)$. Inserting Eq.~\eqref{eq:S-ImQ} of Sec.~\ref{S:spin1} into Eq.~\eqref{eq:S-ReImF}, each of the two antisymmetric contractions equals $\sum_{j\neq k}\sin[2\pi(k-j)/3]\,(d^{(m_j)}q^{(m_k)}-q^{(m_j)}d^{(m_k)})=-9$, so that
\begin{equation}
\operatorname{Im}F(Z_0,Z_x)=\frac{\Lambda}{2}.
\end{equation}
The clock OTOC thus measures the single imaginary coordinate of the spin-1 quasiprobability, but it requires operators that are not Hermitian. The generalized correlators $F'_{abce}=\sum a^{(n)}b^{(m)}c^{(n')}e^{(m')}Q_{nm,n'm'}$, with $a$ and $c$ ranging over a basis of operators diagonal in the levels of site $x$, and $b$ and $e$ over such a basis at site $0$, are related to $Q$ by an invertible linear map, so the family $F'$ carries exactly the information in $Q$. Here the four operators are chosen independently, and $\operatorname{Im}F'$ contains $\operatorname{Im}Q$ already for Hermitian $a$, $b$, $c$ and $e$; for example, $\Lambda=-2\operatorname{Im}F'_{ddqq}$ with $F'_{ddqq}=\Tr[d_x(t)d_0q_x(t)q_0]$.

\section{Reversing the Hamiltonian}
\label{S:reversal}

Let complex conjugation be taken in a product basis in which the level projectors $P^{(n)}$ are real, for example diagonal. Replacing $H$ by $-H^*$ replaces $U=e^{-\ii Ht}$ by $U^*$, so $A_n=U^\dagger P^{(n)}_xU$ becomes $A_n^*$, while $B_m=B_m^*$ is unchanged. The normalized trace of a complex-conjugated product is the complex conjugate of the trace, hence $Q\to Q^*$. The real part of $Q$ and the distribution $p_{nm}$ are unchanged, and the imaginary part changes sign. Since every time-ordered correlation of the two measurements is a moment of $p_{nm}$ (Lemma~\ref{lem:TOC}), and the OTOC of Hermitian observables diagonal in the measured levels contains only $\operatorname{Re}Q$ [Eq.~\eqref{eq:S-ReImF}], none of these distinguishes $H$ from $-H^*$.

\section{The probe algorithm}
\label{S:probe}

The algorithm rests on two identities: a 2-design average that turns $Q$ into covariances of two-time quasiprobabilities, and the probe signals that measure both components of these quasiprobabilities. For pure states $\rho_u$ drawn from a 2-design, $\E_u(\rho_u\otimes\rho_u)=(\id+\mathsf S)/[\mathcal D(\mathcal D+1)]$, where $\mathsf S$ swaps the two copies~\cite{Dankert2009,Elben2023}. Contracting with $X\otimes Y$ for arbitrary, not necessarily Hermitian, operators $X$ and $Y$ gives
\begin{equation}
\Tr(XY)-\Tr X\,\Tr Y=(\mathcal D+1)\Cov_u\big(\langle X\rangle_u,\langle Y\rangle_u\big),
\label{eq:S-random}
\end{equation}
with the unconjugated covariance $\Cov(x,y)=\E(xy)-\E x\,\E y$. With $X=A_nB_m$ and $Y=A_{n'}B_{m'}$ this is Eq.~\eqref{eq:randomstates}. With $Y^\dagger$ in place of $Y$, the trace becomes the time-ordered kernel $\Tr(A_nB_mB_{m'}A_{n'})=\delta_{nn'}\delta_{mm'}p_{nm}$, so the conjugated covariance is real. Write $\alpha=(n,m)$, $\beta=(n',m')$ and $z_{u,\alpha}=a_{u,\alpha}+\ii b_{u,\alpha}$. The two relations then give
\begin{equation}
\begin{aligned}
\Cov(a_\alpha,a_\beta)+\Cov(b_\alpha,b_\beta)&=\frac{\delta_{\alpha\beta}p_\alpha-p_\alpha p_\beta}{\mathcal D+1},\qquad
\Cov(a_\alpha,a_\beta)-\Cov(b_\alpha,b_\beta)=\frac{\operatorname{Re}Q_{\alpha\beta}-p_\alpha p_\beta}{\mathcal D+1},\\
\Cov(a_\alpha,b_\beta)&=\Cov(b_\alpha,a_\beta)=\frac{\operatorname{Im}Q_{\alpha\beta}}{2(\mathcal D+1)},
\end{aligned}
\end{equation}
which contain the relations for $\operatorname{Re}Q$ and $\operatorname{Im}Q$ in the main text. The means are $\E_ua_{u,\alpha}=p_\alpha$ and $\E_ub_{u,\alpha}=0$, and since $\operatorname{Im}Q_{\alpha\alpha}=0$, the real and imaginary parts of a single feature are uncorrelated. The identity requires a 2-design on the full Hilbert space; products of independent random states on single sites do not satisfy it. Each state must be replayed for every setting, so that the features of one label $u$ refer to the same state. They are estimated from independent shots, and shot noise therefore does not bias products of different features; a diagonal entry $Q_{\alpha\alpha}$ requires two independent records of $z_{u,\alpha}$, as for the OTOC estimator in Sec.~\ref{S:spin1}.

\textit{Probe signals.} The probe starts in $\ket0$, the $+1$ eigenstate of $\sigma_z$. In the interaction picture the two couplings are $V_1=e^{-\ii\eta_1g\sigma_\nu\otimes B_m}$ and $V_2=e^{-\ii\eta_2g\sigma_x\otimes A_n}$, with $A_n=U^\dagger P^{(n)}_xU$, and the probe reads out $V_1^\dagger V_2^\dagger\sigma_zV_2V_1$. Expanding each conjugation as $e^{\ii\theta X}Ye^{-\ii\theta X}=Y+\ii\theta[X,Y]+O(\theta^2)$, the term proportional to $\eta_1\eta_2$ is $-\eta_1\eta_2g^2[\sigma_\nu\otimes B_m,[\sigma_x\otimes A_n,\sigma_z]]$. The sign-alternating sum keeps only the terms odd in both signs, so that
\begin{equation}
s_\nu=-g^2\big\langle\big[\sigma_\nu\otimes B_m,[\sigma_x\otimes A_n,\sigma_z]\big]\big\rangle+O(g^4),
\label{eq:S-probe}
\end{equation}
with the expectation value taken in $\ket0\bra0\otimes\rho_u$. Since $[\sigma_x\otimes A_n,\sigma_z]=-2\ii\sigma_y\otimes A_n$, the double commutator equals $2\ii[A_n,B_m]$ for $\nu=y$ and $2\sigma_z\otimes\{A_n,B_m\}$ for $\nu=x$. With $\langle[A_n,B_m]\rangle_u=2\ii\operatorname{Im}z_u$ and $\langle\{A_n,B_m\}\rangle_u=2\operatorname{Re}z_u$, this gives $s_y=4g^2\operatorname{Im}z_u$ and $s_x=-4g^2\operatorname{Re}z_u$. At finite $g$, each setting measures a system operator that can be computed exactly from the two couplings; the resulting bias of the estimates scales as $g^2$. The probe stays close to $\ket0$, so each shot gives $\sigma_z=-1$ with a probability of order $g^2$, and the shot-noise variance of $z_u$ from $N_s$ shots per setting scales as $g^{-2}N_s^{-1}$.

\begin{algorithm}[H]
\caption{Learning $Q$ with a local probe.}
\label{alg:probe}
\begin{algorithmic}[1]
\Require projectors $\{B_m\}$ at site $0$ and $\{P^{(n)}_x\}$ at site $x$; time $t$; coupling $g$; number of states $R$; shots $N_s$
\For{$u=1,\dots,R$}
  \State draw $\rho_u$ from a global 2-design \Comment{replayed for all settings below}
  \For{each pair $(n,m)$}
    \For{$\nu\in\{x,y\}$}
      \For{$\eta_1,\eta_2\in\{+1,-1\}$}
        \State prepare $\ket0\bra0\otimes\rho_u$; apply $e^{-\ii\eta_1g\sigma_\nu\otimes B_m}$, then $U(t)$, then $e^{-\ii\eta_2g\sigma_x\otimes P^{(n)}_x}$
        \State estimate $\langle\sigma_z\rangle_{\eta_1\eta_2}$ from $N_s$ probe shots
      \EndFor
      \State $s_\nu\gets\frac14\sum_{\eta_1\eta_2}\eta_1\eta_2\langle\sigma_z\rangle_{\eta_1\eta_2}$
    \EndFor
    \State $z_{u,nm}\gets(-s_x+\ii s_y)/(4g^2)$
  \EndFor
\EndFor
\State $p_{nm}\gets R^{-1}\sum_u\operatorname{Re}z_{u,nm}$
\State $Q_{nm,n'm'}\gets p_{nm}p_{n'm'}+(\mathcal D+1)\,\widehat{\Cov}_u(z_{u,nm},z_{u,n'm'})$ \Comment{unbiased sample covariance}
\State \Return $Q$
\end{algorithmic}
\end{algorithm}

\textit{Cost.} Each random state is used for $8rs$ settings of $N_s$ shots. Targeted contractions of $Q$ need fewer settings; the spin-1 estimators of Sec.~\ref{S:spin1} use three records per state. Each entry of $Q$ is estimated from a sample covariance of the form $(\mathcal D+1)\Cov_u(x_u,y_u)$, with $x_u=\langle X\rangle_u$ and $y_u=\langle Y\rangle_u$, so its statistical error is the single-state standard deviation divided by $\sqrt R$. For Haar-random states and large $\mathcal D$, the expectation values behave as Gaussian variables with covariances $\Tr(X_cY_c)/(\mathcal D+1)$, where $X_c=X-\Tr X$, and the single-state variance of $(\mathcal D+1)(x_u-\bar x)y_u$ becomes $\Tr(X_c^2)\Tr(Y_c^2)+[\Tr(X_cY_c)]^2$. These normalized traces depend on the probed local operators and on the dynamics, but not on $\mathcal D$, so the number of random states needed for a given relative precision does not grow with system size. For $X=Y$, as for an OTOC, the relative standard deviation is $\sqrt2$ whatever the operator. At the reference point of Fig.~\ref{fig:S-probe}, it is $11.8$ for the estimator of $\Lambda$; sampling reproduces both values to within $3\%$ for $L=5$ to $7$ (Table~\ref{tab:S-finite}). A state 2-design fixes the means of the estimators, whereas their variances depend on fourth moments, so these values refer to Haar-random states. Shot noise scales differently. Over random states a feature fluctuates by an amount of order $\mathcal D^{-1/2}$, whereas its shot noise is of order $(g^2N_s)^{-1/2}$, so the shot-noise part of the per-state variance of the estimators is controlled by $\mathcal D/(g^2N_s)$. Keeping it at a fixed fraction of the spread over states requires $N_s\propto\mathcal D/g^2$, and the total number of preparations then grows in proportion to $\mathcal D$, as in other protocols based on global random states~\cite{Elben2023} (Sec.~\ref{S:errors}).

\textit{Other routes to $Q$.} Sequential weak or strong measurements interleaved with forward and backward evolution reconstruct $Q$~\cite{YungerHalpern2018,Dressel2018,Mohseninia2019}, and interferometric schemes condition the many-body evolution on an ancilla~\cite{Swingle2016,YungerHalpern2018}. Randomized-measurement protocols correlate the expectation value of a late observable on random states with and without an early unitary kick~\cite{Vermersch2019}. For Hermitian observables and kicks this yields contractions of $\operatorname{Re}Q$ only. With kicks of complex spectrum and readout of the many-body system, the same protocols reach all of $Q$. Let $V_{\bm\varphi}=\sum_me^{\ii\varphi_m}B_m$ be an early phase kick. Eq.~\eqref{eq:S-random} with $X=A_{n'}$ and $Y=V_{\bm\varphi}^\dagger A_nV_{\bm\varphi}$, both Hermitian, gives $\Tr(A_{n'}V_{\bm\varphi}^\dagger A_nV_{\bm\varphi})$ from the covariance of two populations of the level $x$: one measured after forward evolution alone, the other after the kick and the same evolution. Expanding the kick
\begin{equation}
\Tr(A_{n'}V_{\bm\varphi}^\dagger A_nV_{\bm\varphi})=\sum_{mm'}e^{\ii(\varphi_{m'}-\varphi_m)}\,Q_{n'm,nm'}.
\label{eq:S-kick}
\end{equation}
For $m\neq m'$ the frequency vectors $\bm e_{m'}-\bm e_m$ are distinct, so a discrete Fourier transform over phases $\varphi_m\in\{0,2\pi/K,\dots,2\pi(K-1)/K\}$ with $K\geq3$ isolates every entry with $m\neq m'$, and the entries with $m=m'$ follow from the one-label sums. One overall phase is irrelevant, so $K^{s-1}$ kick settings suffice, and a projective readout of site $x$ provides all $r$ late outcomes at once. A kick with only two distinct phases, such as a Hermitian unitary, has collinear eigenvalues, and Eq.~\eqref{eq:S-kick} then contains only $\operatorname{Re}Q$, in line with Sec.~\ref{S:ImQ}. This route needs a projective measurement of the many-body system at site $x$, whereas the probe route needs only the probe to be measured and couples to the system through two controlled unitaries. Which route is preferable therefore depends on the platform: where site-resolved readout is available, the kick route uses fewer settings per random state, and where the system can only be addressed through an auxiliary spin~\cite{Randall2021,Meinel2022}, the probe route applies.

\section{The spin-1 chain}
\label{S:spin1}

\textit{Decomposition of $Q$.} For commuting $A,C$ and commuting $B,D$, expanding the products of commutators and anticommutators and using cyclicity gives $\Tr(ABCD)=-\frac12\Tr(M_{AB}M_{CD})-\frac{\ii}{2}\Tr(N_{AB}M_{CD})+\Tr(CABD)$, with $M_{AB}=\ii[A,B]$ and $N_{AB}=\{A,B\}$. For $A=A_n$, $B=B_m$, $C=A_{n'}$ and $D=B_{m'}$ the last term is $\delta_{nn'}\delta_{mm'}p_{nm}$, so
\begin{equation}
Q_{nm,n'm'}=\delta_{nn'}\delta_{mm'}p_{nm}-\tfrac12\Tr(M_{nm}M_{n'm'})-\tfrac{\ii}{2}\Tr(N_{nm}M_{n'm'}),
\label{eq:S-anatomy}
\end{equation}
with $M_{nm}=\ii[A_n,B_m]$ and $N_{nm}=\{A_n,B_m\}$. Apart from the classical term, the real part is minus half the Gram matrix of the level responses, and the imaginary part pairs symmetrized correlations with responses. For a spin one, we resolve the level projectors into the dipole $d=\sqrt{3/2}\,S^z$ and the quadrupole $q=\frac{3}{\sqrt2}[(S^z)^2-\frac23]$, $P^{(m)}=(\id+d^{(m)}d+q^{(m)}q)/3$ with $d^{(m)}=\sqrt{3/2}\,(1,0,-1)$ and $q^{(m)}=(1,-2,1)/\sqrt2$ on the levels $m=+1,0,-1$. Identity components drop out of commutators, so $M_{nm}=\frac19\sum_{ab}a^{(n)}b^{(m)}M_{ab}$ with $M_{ab}=\ii[a_x(t),b_0]$ and $N_{ab}=\{a_x(t),b_0\}$ for $a,b\in\{d,q\}$, and $\operatorname{Re}Q$ is fixed by the overlaps $\Tr(M_{ab}M_{ce})$ of the four responses and by the two-time correlations $\Tr[a_x(t)b_0]$ that enter $p_{nm}$. The squared lengths $\Tr(M_{aa}^2)$ are the squared-commutator OTOCs of $d$ and $q$. Those of $S^z$ and $(S^z)^2$ shown in Fig.~\ref{fig:spin1} are $\frac49\Tr(M_{dd}^2)$ and $\frac4{81}\Tr(M_{qq}^2)$. The imaginary part reduces to
\begin{equation}
\operatorname{Im}Q_{nm,n'm'}=-\frac{\Lambda}{162}\big(d^{(n)}q^{(n')}-q^{(n)}d^{(n')}\big)\big(d^{(m)}q^{(m')}-q^{(m)}d^{(m')}\big),
\label{eq:S-ImQ}
\end{equation}
a single number times a fixed pattern, as required by the bound of one imaginary coordinate for $r=s=3$. Here $\Lambda=\Tr(N_{dd}M_{qq})=\Tr(N_{qq}M_{dd})=-2\operatorname{Im}\Tr[d_x(t)d_0q_x(t)q_0]$, where cyclicity gives the equality of the two contractions. The pattern takes a simple form. For any two different levels, $d^{(n)}q^{(n')}-q^{(n)}d^{(n')}=-\sqrt3\,\sigma_{nn'}$, where $\sigma_{nn'}=+1$ if $n\to n'$ follows the cyclic order $+1\to0\to-1\to+1$ and $\sigma_{nn'}=-1$ otherwise. Every entry in which both outcomes change therefore has
\begin{equation}
\operatorname{Im}Q_{nm,n'm'}=-\frac{\Lambda}{54}\,\sigma_{nn'}\sigma_{mm'},
\label{eq:S-cyclic}
\end{equation}
so that all of them have modulus $|\Lambda|/54$, with the sign set by the orientations of the two transitions. This is the statement $\operatorname{Im}Q=\pm\Lambda/54$ of the main text.

\textit{Symmetries of $\Lambda$.} With separate couplings, $H=J_zH_{zz}+J_\perp H_\perp+\Delta H_\Delta$, two operations constrain $\Lambda$. The Hamiltonian is real, so reversing it reverses $\Lambda$ (Sec.~\ref{S:reversal}). On a bipartite chain, the rotation $\prod_{j\,\mathrm{odd}}e^{\ii\pi S^z_j}$ reverses $J_\perp$ and leaves every diagonal operator unchanged. Together these give $\Lambda(J_z,J_\perp,\Delta;t)=-\Lambda(-J_z,J_\perp,-\Delta;t)$: $\Lambda$ is odd in the diagonal energies and vanishes for pure hopping.

\textit{Short-time expansion of $\Lambda$.} Expanding $a_x(t)=\sum_k(\ii t)^k\operatorname{ad}_H^k(a_x)/k!$, with $\operatorname{ad}_H(X)=[H,X]$, gives $\Lambda=\sum_nc_nt^n$ with
\begin{equation}
c_n=-2\operatorname{Im}\bigg[\ii^n\sum_{k=0}^{n}\frac{\Tr\big[\operatorname{ad}_H^k(d_x)\,d_0\,\operatorname{ad}_H^{n-k}(q_x)\,q_0\big]}{k!\,(n-k)!}\bigg].
\end{equation}
Each $c_n$ is a finite trace of local operators and a homogeneous polynomial of degree $n$ in $(J_z,J_\perp,\Delta)$. Since $H$ is real, $\Lambda$ is odd in $t$. Each term of $\operatorname{ad}_H^k(d_x)$ is a product of bond terms along a connected set of at most $k$ bonds that contains site $x$, and the trace with $d_0$ and $q_0$ requires the union of the two sets to contain site $0$. For $x=1$, $c_n$ is thereby a sum over connected sets of at most $n$ bonds containing sites $0$ and $1$, and a ring of $N$ sites reproduces the infinite chain for $n\leq N-1$, since no such set then closes around the ring. For $x=1$ the odd coefficients, evaluated exactly on a ring of seven sites, are $c_1=c_3=0$ and $c_5=J_\perp^4(J_z-2\Delta)$. The diagonal factor is, up to its sign, the energy difference $E_{+-}-E_{00}=2\Delta-J_z$ between the two pair states that the transverse term connects. The next coefficient, evaluated on a ring of eight sites, is $c_7=-3.822J^7$ at $\Delta=0$ and $-0.900J^7$ at $\Delta=J/2$, for $J_z=J_\perp=J$, so the resonance removes only the leading term.

\textit{Numerics and estimators.} All curves use exact diagonalization of periodic chains with $\mathcal D=3^L$, $L=6$ for Fig.~\ref{fig:spin1} and $L=5$ for Fig.~\ref{fig:S-probe}, with $a_x(t)$ computed in the eigenbasis of $H$. The probe couples to $d$ and $q$ at sites $0$ and $x$; the coupling to $q$ requires a term proportional to $(S^z)^2$. For observable $a$, the records are $s^{(a)}_y=-2g^2\langle M_{aa}\rangle_u$ and $s^{(a)}_x=-2g^2\langle N_{aa}\rangle_u$ to leading order. The OTOC of $S^z$ is estimated as $\frac49(\mathcal D+1)\Cov_u(s^{(d)}_y,s'^{(d)}_y)/(4g^4)$ from two records of the same setting taken with independent shots, so that their shot noise does not bias the product. The coordinate $\Lambda$ is estimated as $(\mathcal D+1)\Cov_u(s^{(q)}_x,s^{(d)}_y)/(4g^4)$; the covariance removes the nonzero mean $\Tr N_{qq}$ of the symmetrized correlation. Of the two equivalent contractions, the pairing of $N_{qq}$ with $M_{dd}$ has the smaller variance here, because $N_{aa}$ carries a large static part and the OTOC of $d$ is smaller than that of $q$. The three records per random state are built from $N_s=10^4$ binary probe outcomes per sign setting, so that each state uses $12N_s$ preparations. At finite $g$, each sign setting measures a system operator computed exactly from the two couplings, which gives the bias in the inset of Fig.~\ref{fig:S-probe}. The points of Fig.~\ref{fig:spin1}(c) use the same estimator of $\Lambda$ for $L=5$ and $Jt=0.5$, with $g=0.15$, $N_s=10^4$ and $R=2\times10^5$ random states at each detuning; their error bars are standard errors, obtained from the spread of the single-state products. Over the scan, the exact curve for $L=6$ differs from that for $L=5$ by less than $10^{-3}$ of its maximum.

\section{Error budget and finite-size checks}
\label{S:errors}

\begin{figure}[!htb]
\centering
\includegraphics{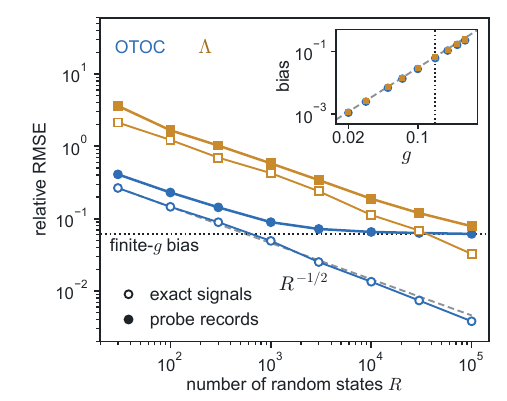}
\caption{Probe reconstruction in the spin-1 chain at the reference point $\Delta=2J$, $x=1$, $Jt=0.7$, $L=5$. Relative root-mean-square error over 100 trials versus the number $R$ of random states, for exact signals and simulated probe records with $g=0.15$ and $10^4$ shots per setting. Dashed, $R^{-1/2}$; dotted, finite-$g$ bias of the OTOC estimate. Both errors fall as $R^{-1/2}$ until the probe records reach the finite-$g$ bias, which more random states cannot remove. Inset: relative bias of the probe estimates of the OTOC of $S^z$ and of $\Lambda$ at finite coupling $g$, from exact finite-$g$ means; here dashed: $3g^2$, while dotted: $g=0.15$.}
\label{fig:S-probe}
\end{figure}

\textit{Sources of error.} Three sources contribute to the error of the probe estimates in Fig.~\ref{fig:S-probe}. First, at finite $g$ each sign setting measures a system operator that is computed exactly from the two couplings, and the estimates converge to finite-$g$ values that lie below the exact ones by about $3g^2$; at $g=0.15$ the bias is $b=-6.10\%$ for the OTOC and $-6.55\%$ for $\Lambda$. Second, the records fluctuate over random states, which sets the per-state standard deviation $\sigma_{\rm st}$ of each estimator. Third, each record is the alternating sum of four probe averages, each estimated from $N_s$ binary outcomes. Given the state, its shot-noise variance is $v=\frac1{16}\sum_{\eta_1\eta_2}(1-m_{\eta_1\eta_2}^2)/N_s$, where $m_{\eta_1\eta_2}$ is the mean of $\sigma_z$ in that setting. The probe stays close to $\ket0$, so $1-m_{\eta_1\eta_2}^2$ is of order $g^2$ and $v$ of order $g^2/N_s$. Records from different settings, or from independent shots of the same setting, are independent, so shot noise does not bias the products of records, but it adds to the per-state variance. For the single-state estimator $c\,(s-\bar s)(s'-\bar s)$ of the OTOC, with $c=\frac49(\mathcal D+1)/(4g^4)$ and $s,s'$ two independent records of $s^{(d)}_y$, the added variance is
\begin{equation}
\sigma_{\rm sh}^2=c^2\,\E_u\Big[2\big(s^{(d)}_y-\bar s\big)^2v+v^2\Big],
\label{eq:S-shot}
\end{equation}
and for $\Lambda$ the analogous expression contains the two different records $s^{(q)}_x$ and $s^{(d)}_y$, each taken relative to its mean over states.

\textit{Budget at the reference point.} Table~\ref{tab:S-budget} lists the three contributions for the parameters of Fig.~\ref{fig:S-probe}, with $\sigma_{\rm st}$ and $\sigma_{\rm sh}$ obtained by averaging over $6\times10^4$ Haar-random states. The predicted root-mean-square error for $R$ states, $[b^2+(\sigma_{\rm st}^2+\sigma_{\rm sh}^2)/R]^{1/2}$, agrees with the simulated runs of Fig.~\ref{fig:S-probe} to within $11\%$, comparable to the statistical uncertainty of about $7\%$ of a root-mean-square error estimated from $100$ runs. For the OTOC, the bias dominates already at $R\approx3\times10^3$. For $\Lambda$, shot noise and the spread over states contribute comparably, and at $R=10^5$ the bias is still the largest single contribution.

\begin{table}[!htb]
\centering
\caption{Error budget of the probe estimates at the reference point of Fig.~\ref{fig:S-probe} ($L=5$, $x=1$, $\Delta=2J$, $Jt=0.7$, $g=0.15$, $N_s=10^4$). Standard deviations refer to a single random state and, like the errors, are given relative to the exact values. Predicted errors use the budget; simulated errors are from the $100$ runs of Fig.~\ref{fig:S-probe}.}
\label{tab:S-budget}
\begin{tabular}{lcc}
\toprule
& OTOC of $S^z$ & $\Lambda$\\
\midrule
exact value & $0.2348$ & $-0.07422$\\
finite-$g$ bias $b$ & $-6.10\%$ & $-6.55\%$\\
\addlinespace
\multicolumn{3}{l}{single-state relative standard deviation}\\
\quad exact signals ($g\to0$, no shot noise) & $1.42$ & $11.7$\\
\quad spread over random states, $\sigma_{\rm st}$ & $1.33$ & $10.9$\\
\quad shot noise, $\sigma_{\rm sh}$ & $1.64$ & $14.6$\\
\addlinespace
\multicolumn{3}{l}{relative RMSE, predicted / simulated}\\
\quad $R=10^4$ & $0.065$ / $0.066$ & $0.193$ / $0.188$\\
\quad $R=3\times10^4$ & $0.062$ / $0.063$ & $0.124$ / $0.120$\\
\quad $R=10^5$ & $0.061$ / $0.062$ & $0.087$ / $0.079$\\
\bottomrule
\end{tabular}
\end{table}

\begin{figure}[!htb]
\centering
\includegraphics{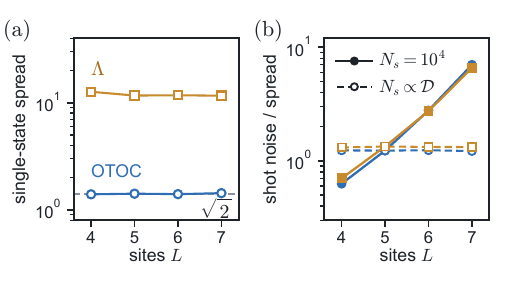}
\caption{Sample complexity of the probe estimators in the spin-1 chain ($x=1$, $\Delta=2J$, $Jt=0.7$, $g=0.15$, $\mathcal D=3^L$). (a)~Single-state relative standard deviation of the estimates of the OTOC of $S^z$ and of $\Lambda$ for exact signals, which sets the number of random states, $R\sim\sigma^2/\epsilon^2$; dashed, $\sqrt2$. (b)~Shot noise relative to the spread over random states, for $N_s=10^4$ shots per setting (filled) and for $N_s=10^4\mathcal D/243$ (open).}
\label{fig:S-scaling}
\end{figure}

\textit{Dependence on system size.} Over random states each record fluctuates by an amount of order $g^2\mathcal D^{-1/2}$, while its shot noise is of order $g\,N_s^{-1/2}$. Relative to $\sigma_{\rm st}^2$, the terms of $\sigma_{\rm sh}^2$ that contain one record are therefore of order $\mathcal D/(g^2N_s)$, and the last term of Eq.~\eqref{eq:S-shot} of order $[\mathcal D/(g^2N_s)]^2$; at the reference point $\mathcal D/(g^2N_s)=1.1$. Table~\ref{tab:S-finite} shows both behaviours. The exact quantities and the spread over states change little with $L$: from $L=5$ on, the exact values agree with those for $L=7$ to within $0.3\%$ and the spreads to within $2\%$, comparable to their sampling error. At fixed $N_s$, in contrast, the shot noise grows with $\mathcal D$. Scaling $N_s$ with $\mathcal D$ compensates this growth: with $N_s=10^4\mathcal D/243$, $\sigma_{\rm sh}$ stays between $1.62$ and $1.64$ for the OTOC and between $14.3$ and $15.5$ for $\Lambda$ from $L=4$ to $7$ [Fig.~\ref{fig:S-scaling}(b)]. The number of random states needed for a given precision therefore does not grow with system size, but the number of shots per setting does, and the total number of preparations, $12N_sR$ for the spin-1 estimators, grows in proportion to $\mathcal D$, as in other protocols based on global random states~\cite{Elben2023}.

\begin{table}[!htb]
\centering
\caption{Finite-size data at the reference point of Fig.~\ref{fig:S-probe} ($x=1$, $\Delta=2J$, $Jt=0.7$) for periodic chains of $L$ sites. Single-state relative standard deviations from sampling over Haar-random states, for exact signals and for the shot noise at $g=0.15$ and $N_s=10^4$.}
\label{tab:S-finite}
\begin{tabular}{rrccccccc}
\toprule
& & \multicolumn{2}{c}{exact value} & \multicolumn{2}{c}{exact signals} & \multicolumn{2}{c}{shot noise}\\
\cmidrule(lr){3-4}\cmidrule(lr){5-6}\cmidrule(lr){7-8}
$L$ & $\mathcal D$ & OTOC & $\Lambda$ & OTOC & $\Lambda$ & OTOC & $\Lambda$\\
\midrule
$4$ & $81$ & $0.2395$ & $-0.07075$ & $1.40$ & $12.7$ & $0.83$ & $8.3$\\
$5$ & $243$ & $0.2348$ & $-0.07422$ & $1.42$ & $11.7$ & $1.64$ & $14.6$\\
$6$ & $729$ & $0.2345$ & $-0.07441$ & $1.40$ & $11.7$ & $3.66$ & $30.0$\\
$7$ & $2187$ & $0.2345$ & $-0.07442$ & $1.43$ & $11.6$ & $9.4$ & $71$\\
\bottomrule
\end{tabular}
\end{table}

\textit{Sample complexity.} For a target relative error $\epsilon$, require the bias, about $3g^2$, and the statistical error, $\sigma/\sqrt R$, each to be at most $\epsilon/\sqrt2$, where $\sigma^2=\sigma_{\rm st}^2+\sigma_{\rm sh}^2$. This gives $g^2\approx\epsilon/(3\sqrt2)$ and $R=2\sigma^2/\epsilon^2$. With $N_s=\kappa\mathcal D/g^2$, the ratio $\sigma_{\rm sh}/\sigma_{\rm st}$ depends on $\kappa$ but not on $\mathcal D$, so $\sigma$ and $R$ are independent of system size, and the total number of preparations is $n_{\rm set}N_sR\approx6\sqrt2\,\kappa\,n_{\rm set}\mathcal D\sigma^2/\epsilon^3$, with $n_{\rm set}=8rs$ settings for the full $Q$ and $n_{\rm set}=12$ for the two spin-1 estimators. The OTOC estimator has $\sigma_{\rm st}\approx\sqrt2$, whereas $\Lambda$ has $\sigma_{\rm st}\approx11$ at the reference point, so that $\Lambda$ needs about $70$ times more random states for the same relative precision.

\textit{Choice of the reference point.} The protocol is the same for any distance $x$ and time $t$, but the number of random states needed for a given relative precision scales as the square of the single-state standard deviation divided by the estimated value. For $\Lambda$, which starts at fifth order in $t$ for $x=1$, this ratio is smallest where $|\Lambda|$ is large. At $x=1$ and $\Delta=2J$ ($L=6$, exact signals), the relative standard deviation of the estimator of $\Lambda$ is $21.0$, $13.7$, $11.8$, $13.3$ and $20.7$ at $Jt=0.5$, $0.6$, $0.7$, $0.8$ and $0.9$, and $|\Lambda|$ reaches its first extremum near $Jt=0.75$. At $x=2$, which the operator front reaches later, it is about $1400$ at $Jt=0.7$ and $45$ at $Jt=1.5$, so that a relative precision of $10\%$ requires about $2\times10^5$ random states, compared with $1.4\times10^4$ at the reference point. The OTOC estimator, with its operator-independent spread $\sqrt2$, is insensitive to this choice.

\textit{Finite size of the curves.} The curves of Fig.~\ref{fig:spin1}, computed for $L=6$, differ from those for $L=7$ by at most $0.6\%$ of their maximum over the time range of Fig.~\ref{fig:spin1}(b), the largest difference occurring for $\Lambda$ at $\Delta=0$, and by less than $10^{-5}$ of their maximum over the scan of Fig.~\ref{fig:spin1}(c).

\section{Beyond infinite temperature}
\label{S:finiteT}

For an initial state $\rho$ other than the identity, define $Q^\rho_{nm,n'm'}=\tr(\rho A_nB_mA_{n'}B_{m'})$ and the time-ordered kernel $K^\rho_{nm,n'm'}=\tr(\rho B_mA_nA_{n'}B_{m'})$, with $\tr\rho=1$. Table~\ref{tab:S-finiteT} lists which statements of the main text survive; the arguments follow.

\begin{table}[!htb]
\centering
\caption{Working boundary of the results. Rows: initial states; ``commutes with $A$'' means $[\rho,A_n]=0$ for all $n$, with $A_n$ in the Heisenberg picture, and likewise for $B$. The equivalence of negativity and incompatibility requires a full-rank $\rho$. A thermal state $\rho\propto e^{-\beta H}$ at finite $\beta$ generically commutes with neither family.}
\label{tab:S-finiteT}
\begin{tabular}{@{}lcccc@{}}
\toprule
& $K^\rho$ is a & negativity $\Leftrightarrow$ & $Q^\rho$ real & $Q^\rho$ real\\
initial state & probability & incompatibility & if $s=2$ & if $r=2$\\
\midrule
$\rho\propto\id$ (infinite temperature) & yes & yes & yes & yes\\
commutes with $A$ and $B$ & yes & yes & yes & yes\\
commutes with $B$ only & yes & yes & no & yes\\
generic, e.g.\ thermal & no & yes & no & no\\
\bottomrule
\end{tabular}
\end{table}

\textit{Negativity and incompatibility for any state.} The three sums used in Lemma~\ref{lem:neg} rely only on the completeness of the two families, so for any $\rho$ the sum of all entries, the sum over entries with $n=n'$ and the sum over entries with $m=m'$ each equal one: for example, $\sum_{nmm'}Q^\rho_{nm,nm'}=\sum_{nm}\tr(\rho A_nB_mA_n)=\sum_n\tr(\rho A_n)=1$. By inclusion and exclusion, the entries in which both labels change sum to $S^\rho-1$, with $S^\rho=\sum_{nm}Q^\rho_{nm,nm}$. With the state-weighted norm $\|X\|_\rho^2=\tr(\rho X^\dagger X)$, expanding the commutator gives $\|[A_n,B_m]\|_\rho^2=\tr(\rho B_mA_nB_m)+\tr(\rho A_nB_mA_n)-2\operatorname{Re}Q^\rho_{nm,nm}$, because $\tr(\rho B_mA_nB_mA_n)$ is the complex conjugate of $Q^\rho_{nm,nm}$. Summing over $n$ and $m$
\begin{equation}
\sum_{n\neq n',\,m\neq m'}\operatorname{Re}Q^\rho_{nm,n'm'}=-\frac12\sum_{nm}\big\|[A_n,B_m]\big\|_\rho^2 ,
\label{eq:S-sumrule-rho}
\end{equation}
which reduces to Eq.~\eqref{eq:sumrule} for $\rho\propto\id$. If all commutators vanish, $Q^\rho_{nm,n'm'}=\delta_{nn'}\delta_{mm'}\tr(\rho A_nB_m)$, where $A_nB_m$ is a projector, so $Q^\rho$ is a probability distribution. For a full-rank state, $\|[A_n,B_m]\|_\rho=0$ only if $[A_n,B_m]=0$. Item~(ii) of Proposition \ref{prop:prop1} hence holds for every full-rank state, thermal states included.

\textit{Time-ordered kernels.} Adjacent late projectors give $K^\rho_{nm,n'm'}=\delta_{nn'}\tr(\rho B_mA_nB_{m'})$. If $\rho$ commutes with every $B_m$, this is $\delta_{nn'}\delta_{mm'}\tr(B_m\rho B_mA_n)\geq0$, a probability, as noted in \cite{YungerHalpern2018}; the same holds for kernels with any number of insertions on a single contour. Otherwise the entries with $m\neq m'$ are the coherences of $\rho$ between eigenspaces of the early measurement, and $K^\rho$ is a quasiprobability of the state itself.

\textit{Two-outcome reality.} If $\rho$ commutes with every $A_n$, then $Q^\rho_{nm,n'm'}=\tr[(X_m\rho^{1/2})^\dagger X_{m'}\rho^{1/2}]$ with $X_m=A_{n'}B_mA_n$, and the closure $\sum_mX_m=0$ gives a real $Q^\rho$ for $s=2$, as in the proof of Theorem~1. If $\rho$ commutes with every $B_m$, the amplitudes $Y_n=B_mA_nB_{m'}$ give a real $Q^\rho$ for $r=2$ in the same way. A generic thermal state commutes with neither family, and its coherences then contribute to $\operatorname{Im}Q^\rho$ even for two outcomes. In the spin-1 chain at $\Delta=2J$, $x=1$, $Jt=0.7$ and $\beta J=1$ ($L=4$), for example, $\operatorname{Im}Q^\rho$ reaches $0.035$ for three outcomes and $0.015$ when both measurements are coarse-grained to the two outcomes $\{+1\}$ and $\{0,-1\}$, and $K^\rho$ has entries with real parts down to $-0.04$.

\textit{Reversal and learning.} The map $H\to-H^*$ of Sec.~\ref{S:reversal} also changes a thermal state $e^{-\beta H}\to e^{\beta H^*}$ so the statement that only $\operatorname{Im}Q$ distinguishes $H$ from $-H^*$ refers to states left invariant by this map like $\rho\propto\id$. The random-state identity~\eqref{eq:S-random} is likewise an infinite-temperature identity; how the probe protocol extends to thermal states is open.

\end{document}